\documentclass[]{aa} 
\usepackage{graphicx}   
\usepackage{natbib}   
\usepackage{color}   
\bibpunct{(}{)}{;}{a}{}{,} 

\newcommand{\CII}{[\ion{C}{ii}]}   
   
\newcommand{\CI}{[\ion{C}{i}]}

\newcommand{\HII}{\ion{H}{ii}}   
\newcommand{\HI}{\ion{H}{i}}

\newcommand{\lsun}{L$_{\odot}$}   
\newcommand{\msun}{M$_{\odot}$}

\newcommand{\degree}{\ensuremath{^\circ}}
\begin{document}   

\title{$^{12}$CO 4--3 and $[$CI$]$ 1--0 at the centers of NGC\,4945 and Circinus} 

   \author{   
     M.\,Hitschfeld\inst{1} \and   
     M.\,Aravena\inst{3} \and    
     C.\,Kramer\inst{1} \and   
     F.\,Bertoldi\inst{3} \and    
     J.\,Stutzki\inst{1} \and
     F.\,Bensch\inst{3} \and
     L.\,Bronfman\inst{4} \and
     M.\,Cubick\inst{1} \and
     M.\,Fujishita\inst{5} \and
     Y.\,Fukui\inst{5} \and
     U.U.\,Graf\inst{1} \and
     N.\,Honingh\inst{1} \and
     S.\,Ito\inst{5} \and
     H.\,Jakob\inst{1} \and
     K.\,Jacobs\inst{1} \and
     U.\,Klein\inst{3} \and
     B.-C.\,Koo\inst{6} \and
     J.\,May\inst{4} \and
     M.\,Miller\inst{1} \and
     Y.\,Miyamoto\inst{5} \and
     N.\,Mizuno\inst{5} \and
     T.\,Onishi\inst{5} \and
     Y.-S.\,Park\inst{6} \and
     J.L.\,Pineda\inst{3} \and
     D.\,Rabanus\inst{1} \and
     M.\,R\"ollig\inst{3} \and
     H.\,Sasago\inst{5} \and
     R.\,Schieder\inst{1} \and
     R.\,Simon\inst{1} \and
     K.\,Sun\inst{1} \and
     N.\,Volgenau\inst{1} \and
     H.\,Yamamoto\inst{5} \and
     Y.\,Yonekura\inst{2} 
          }

   \institute{
     KOSMA, I. Physikalisches Institut, Universit\"at zu K\"oln,
     Z\"ulpicher Stra\ss{}e 77, D-50937 K\"oln, Germany
     \and
     Department of Physical Science, Osaka Prefecture University,
     Osaka 599-8531, Japan
     \and
     Argelander-Institut f\"ur Astronomie,  Auf dem H\"ugel 71,
     D-53121 Bonn, Germany
     \and
     Departamento de Astronom\'{i}a, Universidad de Chile, Casilla 36-D, Santiago, Chile
     \and
     Department of Astrophysics, Nagoya University, Chikusa-ku, Nagoya 464-8602, Japan
     \and
     Seoul National University, Seoul 151-742, Korea
   }

   \offprints{M.\,Hitschfeld, \email{hitschfeld@ph1.uni-koeln.de}}   
   \date{Received / Accepted }   
      
   \abstract
   { Studying molecular gas in the central regions
   of the star burst galaxies NGC\,4945 and Circinus enables us to
   characterize the physical conditions and compare them to previous local and high-z studies.}
   {We estimate temperature, molecular density and column densities of
   CO and atomic carbon. Using model predictions we give a range of estimated CO/C
   abundance ratios.}
   {Using the new NANTEN2 4m sub-millimeter telescope in Pampa La Bola, Chile, we observed
    for the first time CO 4--3 and \CI\ $^3P_1-^3P_0$ at the centers of both
    galaxies at linear scale of 682 pc and 732 pc respectively. We compute the
    cooling curves of $^{12}$CO and $^{13}$CO using radiative transfer models
    and estimate the physical conditions of CO and $[$CI$]$.}
   {The centers of NGC\,4945 and Circinus are very \CI\ bright objects,
    exhibiting \CI\ $^3P_1-^3P_0$ luminosities of 91 and
    67\,Kkms$^{-1}$kpc$^{2}$, respectively. The \CI\ $^3P_1-^3P_0$/CO 4--3 ratio of
    integrated intensities are large at 1.2 in NGC\,4945 and 2.8 in Circinus.
    Combining previous CO $J$= 1--0 ,
    2--1 and 3--2 and $^{13}$CO $J$= 1--0 , 2--1  studies with our new observations, the
    radiative transfer calculations give a range of densities, $n(\rm H_{2})=10^{3}-3 \times 10^{4}$cm$^{-3}$, and a wide range of  kinetic
   temperatures, $T_{\rm kin}= 20-100$K, depending on the density. To discuss the degeneracy in density and
   temperature, we study two representative solutions. In both galaxies the
   estimated total $[$CI$]$ cooling intensity is stronger by factors of $\sim 1-3$
   compared to the total CO cooling intensity. The CO/C abundance ratios are 0.2-2,
    similar to values found in Galactic translucent clouds.   
     }
   {Our new observations enable us to further constrain the excitation
   conditions and estimate the line emission of higher--$J$ CO-- and the upper $[$CI$]$--lines. For the
   first time we give estimates for the CO/C abundance ratio in the center regions
   of these galaxies. Future CO $J$= 7--6 and $[$CI$]$ 2--1 observations will
   be important to resolve the ambiguity in the physical conditions and
   confirm the model predictions.}
  
   \authorrunning{Hitschfeld et al.} 
   \titlerunning{CO 4--3 and $[$CI$]$ 1--0 in the centers of NGC\,4945 and Circinus}   
   \maketitle   
   
\section{Introduction} 

The spiral galaxies NGC\,4945 and Circinus at  distances of 
$\sim 3.7$ and $\sim 4$\,Mpc belong to the nearest and infrared brightest galaxies in
the sky.  Their strong central star burst activity is fed by large
amounts of molecular material and this has been studied extensively at
millimeter wavelengths \citep[e.g.][]{Curran2001,Wang2004}.  However,
sub-millimeter observations are largely missing. The important
rotational transitions of CO with $J\ge4$ and the fine structure
transitions of atomic carbon have not yet been observed. These
transitions often contribute significantly to the thermal budget of
the interstellar gas in galactic nuclei and are therefore important
tracers of the physical conditions of the warm and dense gas. 

Emission of CO, \CI\ (and \CII) traces the bulk of carbon bearing
species in molecular clouds that play an important role in their
chemical network. The CO 4--3 transition in particular is a sensitive
diagnostic of the dense and warm gas while the CO 1--0 transition traces
the total molecular mass. The \CI\ $^3P_1-^3P_0$ (henceforth 1--0)
line has, to date, been detected in about 30 galactic nuclei
\citep{Gerin2000,IsraelBaas2002,Israel2004}
and appears to trace the surface regions of clumps
illuminated by the FUV field of newborn, massive stars
\citep[e.g.][]{Kramer2004, Kramer2005}. The use of CI as an accurate tracer of
the cloud mass has been discussed with controversy \citep{Frerking1989,Papadopoulos2004, Mookerjea2006}.

The strong cooling emission is balanced by equally strong heating
caused by the vigorous star formation activity in the galaxy centers.
The variation of CO cooling intensities with rotational number, i.e. the
peak of the CO cooling curve, reflects the star forming activity
\citep{bayet2006} and, possibly, also the underlying heating
mechanisms. Several mechanisms have been proposed to explain the
heating of the ISM in galactic nuclei and it is currently rather
unclear which of these dominates in individual sources
\citep[e.g.][]{Wang2004}.  Heating by X-rays from the active galactic
nuclei (AGN) may lead to strongly enhanced intensities of high-$J$ CO
transitions \citep{Meijerink2007}, possibly allowing one to discriminate
this heating mechanism from e.g.  stellar ultraviolet heating via
photon dominated regions (PDRs) \citep[e.g.][]{bayet2006}. The
greatly enhanced supernova rate by several orders of magnitude
relative to the solar system value leads to an enhanced cosmic ray
flux, providing another source of gas heating in the centers
\citep{Farquhar1994,bradford2003}.
Another mechanism for heating is provided by shocks. The most
  important, large-scale shocks are produced by density wave instabilities
  which induce gravitational torques (in spiral arms and/or bars) and make the
 gas fall into the nucleus \citep{Usero2006}. A non-negligible contribution is
 also given by shocks produced by supernovae explosions. On smaller scales,
 bipolar outflows from young stellar objecs (YSO) can also contribute to this
 heating, although to a lesser extent \citep[e.g.][]{garcia-burillo2001}.
 

\subsection{NGC\,4945} 

NGC\,4945, a member of the Centaurus group of galaxies, is seen nearly
edge-on (Table\,\ref{tab_properties}) with an optical diameter of
$\sim 20'$\, \citep{devaucouleurs1991}. \HI\ kinematics indicate a
galaxy mass of $1.4\,10^{11}$\,\msun\ within a radius of $6.3'$ with
molecular and neutral atomic gas contributing $\sim2\%$ respectively
\citep{Ott2001}.

With a dynamical mass of $\sim3\,10^9\,$\msun\ in the central
600\,pc \citep{Mauersberger1996}, it is one of the strongest IRAS
point sources with almost all the far infrared luminosity coming from the nucleus
\citep{Brock1998}. Observations of the X-ray spectrum are
consistent with a Seyfert nucleus \citep{Iwasawa1993}, and further analysis
of optical imaging and infrared spectra \citep{Moorwood1994}
suggests that this object is in a late stage on the transition from starburst
to a Seyfert galaxy. Its nucleus was the first source in which 
a powerful H$_2$O mega maser was detected \citep{DosSantos1979}.

Studies in \HI\ \citep{Ables1987} and low-$J$ CO transitions
\citep{Whiteoak1990,Dahlem1993, Ott1995, Mauersberger1996} suggest the
presence of a face-on circumnuclear molecular ring. Millimeter
molecular multiple transition studies \citep{Wang2004, Cunningham2005}
are consistent with this result.  The bright infrared 
and radio emission in the nucleus \citep{Ghosh1992},
and the evidence that large amounts of 
gas seem to coexist in the central $30''$ \citep{Henkel1994} make it
particularly suited for studying the high density environment in the
center of this galaxy.

\subsection{Circinus}

The nearby starburst spiral Circinus has a small optical angular
diameter of $\sim 7$\arcmin compared to its hydrogen diameter $D_{\rm
  H} = 36'$ defined by the $1\, 10^{20}$atoms cm$^{-2}$ contour in \citet{Freeman1977}.

Circinus has a dynamical mass of $\sim 3\ 10^9\ $ M$_{\odot}$ within the inner
560\,pc \citep{Curran1998} matching the value obtained for NGC\,4945.
Large amounts of molecular gas have been found by studies of low-$J$
CO observations \citep{Johansson1991,Aalto1995,Elmouttie1998,Curran1998} including
C$^{17}$O, C$^{18}$O and HCN \citep{Curran2001}.
The strong H$_{2}$O maser emission found \citep{Gardner1982}
traces a thin accretion disk of 0.8 pc radius, with a significant population of
masers lying away from this disk, possibly in an outflow \citep{Greenhill1998}.
Its obscured nucleus is classified as an X-ray Compton thick Seyfert 2. It shows 
a circumnuclear star burst on scales of 100 - 200\,pc \citep{Maiolino1998}
with a complex structure of \HII\ as seen in Hubble
Space Telescope (HST) H$\alpha$ images \citep{Wilson2000}. Adaptive
optics studies find there has been a recent star burst ($\sim100$ Myr
old) in the central 8\,pc accounting for 2\% of the total luminosity
\citep{MuellerS}. The strong FIR emission, the large molecular gas
reservoir, the existence of a molecular ring associated with star
burst activity \citep{Curran1998} and the similarity with NGC\,4945
make them ideal objects for comparative studies of
the dense and warm ISM in their nuclei.
\section{Observations}   

\begin{center}   
\begin{table}[h*]   
  \caption[]{\label{tab_properties}   
    {\small Basic properties of Circinus, NGC\,4945 and
NGC\,253.
$L_{\rm IR}$ is the total mid and far infrared luminosity
calculated from the flux densities at 12$\mu$m, 25$\mu$m,
60$\mu$m, 100$\mu$m listed in the IRAS point source catalogue
\citep{1985IRAS} using
the formulae given in Table 1 of \citet{Sanders1996} and the distances listed below.
References: $^a$ RC3 catalogue of \citet{devaucouleurs1991}, %
 $^b$
\citet{Mauersberger1996}, $^c$ \citet{Freeman1977}, $^d$
\citet{Curran1998}, $^e$ \citet{Rekola2005}, $^f$
\citet{CurranLFIR2001}, $^g$  \citet{Fullmer1989} . 
 }}
\begin{tabular}{lrrrrr}   
\hline \hline   
                             & Circinus & NGC\,4945 & NGC\,253 \\    
\noalign{\smallskip} \hline \noalign{\smallskip}    
RA(2000)                     & 14:13:09.9  & 13:05:27.4 & 00:47:33.1   \\   
DEC(2000)                    & -65:20:21   & -49:28:05 & -25:17:18\\   
Type                         & SA(s)b$^{a}$ & SB(s)cd$^{a}$& SAB(s)c$^{a}$ \\      
Distance [Mpc]               &$4.0^{d}$ &$3.7^{b}$& 3.5$^{e}$ \\             
$38''$ correspond to         & 732\,pc & 682\,pc & 646\,pc \\                  
LSR velocity [kms$^{-1}$]    & 434 & 555 & 243 \\       
Inclination [deg]            & 65$^{c}$  & $78^{b}$ & $78^{a}$  & \\               
$L_{\rm IR}$ [$10^{10}$ \lsun]            & 1.41$^{g}$  & $1.39^{g}$ & 2.67$^{g}$ & \\      
$S_{100} [{\rm Jy}]$ & $3.16\,10^2$ & $6.86\,10^2$ &  $10.4\,10^2$ \\
%
\noalign{\smallskip} \hline \noalign{\smallskip}   
\end{tabular}   
\end{table}   
\end{center}   
We used the new NANTEN2 sub-millimeter telescope \citep[e.g.][]{Kramer2007} on Pampa
la Bola at an elevation of 4900m together with a dual channel 490/810 GHz
receiver (operated jointly by the Nagoya University radioastronomy group
and the KOSMA group from Universit\"at zu K\"oln) to observe the centers
of NGC\,4945, Circinus, and NGC\,253 (Table \ref{tab_properties}) in CO
4--3 and \CI\ 1--0. The observations were performed from September to
October 2006 using the position-switch mode with 20\,sec On- and
20\,sec Off-time. In October 2007 we reobserved $^{12}$CO 4--3
  in the centers of NGC4945 and
Circinus with the newly available chopping tertiary in double beam-switch mode yielding
significantly improved baselines and removing atmospheric features in the spectra. In
NGC\,4945 the total integration times ON-source are
17\,min and 10\,min for $^{12}$CO 4--3 and \CI\ 1--0 respectively
and in Circinus integration times ON-source are 10\,min and 34\,min for
$^{12}$CO 4--3 and \CI\ 1--0 respectively
at source elevations of 40-60\degree. The relative
calibration uncertainty derived from repeated pointings on the nuclei is about 15\%.

Position-switching was conducted by moving the telescope $10'$ in
azimuth, i.e. out of the galaxy. Using double beam-switch
  (dbs) mode, the chopper throw is fixed at $162''$ in azimuth with a chopping frequency
of 1 Hz.
Typical double-sideband receiver temperatures of the dual-channel 460/810\,GHz
receiver were $\sim$250\,K at 460\,GHz and 492\,GHz. The
system temperature varied between $\sim$850 and $\sim$1200\,K.
%
%
As backends we used two acusto optical spectrometers (AOS) with a
bandwidth of 1\,GHz and a channel resolution of 0.37\,kms$^{-1}$ at
460\,GHz and 0.21\,kms$^{-1}$ at 806\,GHz.
The pointing was regularly checked and pointing accuracy was stable
with corrections of $\sim10''$.  The half power beam width (HPBW) at
460\,GHz and 492\,GHz is 38$''$ with a beam efficiency $B_{\rm eff}=0.50$ and a
forward efficiency $F_{\rm eff}=0.86$ \citep{Simon2007}. The
calibrated data on $T_{A}^{*}$-scale were converted to $T_{\rm mb}$ by
multiplying by the ratio $F_{\rm eff}$/$B_{\rm eff}$.
%
%
The standard calibration procedure derives the atmospheric
transmission (averaged over the bandpass) from the observed
difference spectrum of hot load and blank sky, the latter taken
at a reference position.Next, the model atmosphere {\tt atm} is used to derive the atmospheric
opacity taking into account the sideband imbalance.
This is an important correction, especially when observing the CO 4-3 and [CI] 1-0 lines.
This pipeline produces the standard spectra on the antenna temperature scale ($T_A^*$).
%
%
In addition, we removed baselines up to first order. 

The 810\,GHz channel was not used for these observations due to insufficient
baseline stability.

All data presented in this paper are on the $T_{\rm mb}$ scale. 

\subsection{$^{12}$CO 4--3 in NGC\,253} 
\begin{figure}[h]   
  \centering   
  \includegraphics[angle=-90,width=8cm]{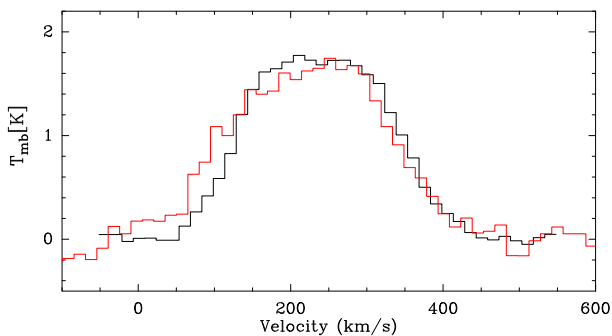}   
\caption{Observed $^{12}$CO 4--3 line emission in NGC\,253 
  from APEX (black line) and NANTEN2 (red line).We attribute small but
  systematic deviations of peak temperatures in the blue wing of the $^{12}$CO
  4--3 line to small pointing variations.}
\label{fig_ngc253}   
\end{figure}

To check 
the NANTEN2 calibration scheme,
we retrieved the $^{12}$CO 4--3 map of NGC\,253 taken at APEX
\citep{Guesten2006} and smoothed it to the NANTEN2 HPBW of $38''$
using a Gaussian kernel. The resulting spectra at the center position are shown in
Fig.\,\ref{fig_ngc253}. Line temperatures and shapes show very good
agreement.


\section{Spectra of CO 4--3 and \CI\ 1--0} 

\begin{figure}[h]   
  \centering   
  \includegraphics[angle=-90,width=8cm]{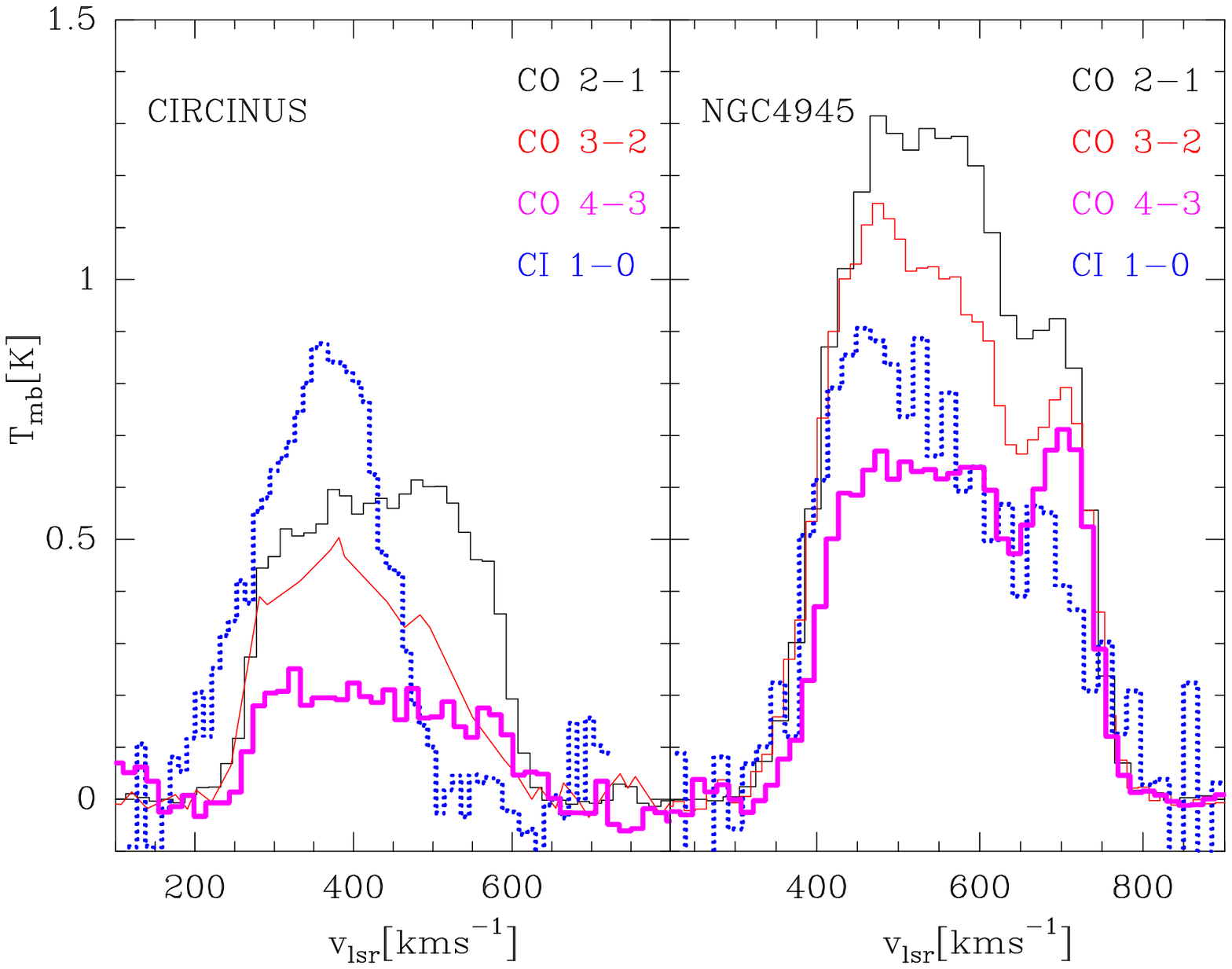}   
  \caption{CO 4--3 and \CI\ 1--0 spectra of Circinus and NGC\,4945
    obtained with the NANTEN2 telescope together with low-$J$ CO
    observations by \citet{Curran2001,Curran1998,Mauersberger1996}.
    All data are at $38''$ resolution, only the CO 3--2 spectrum of
    Circinus is at $15''$ resolution. The small amount
  of excess emission in the blue wing of the \CI\ 1--0 
  line in Circinus maybe caused by small pointing variations in the eastern
  direction as emission in lower-$J$ CO lines extends to this region and velocities \citep{Curran1998}.}
  \label{fig_spectra}
\end{figure}

Figure\,\ref{fig_spectra} shows the $^{12}$CO 4--3 and \CI\ 1--0
spectra of NGC\,4945 and Circinus obtained with the NANTEN2 telescope.
CO 4--3 spectra peak at 700\,mK in NGC\,4945 and 250\,mK in Circinus.
Outside the velocity ranges of 350--800\,kms$^{-1}$ and
200--600\,kms$^{-1}$ respectively, the baseline rms values are 11\,mK 
and 25\,mK, respectively, at the velocity resolution of
15\,kms$^{-1}$.
\CI\ 1--0 spectra peak at 900\,mK in both galaxies while the rms values are
110\,mK and 140\,mK, respectively.  The \CI\ 1--0 area-integrated luminosities are
91\,Kkms$^{-1}$kpc$^{2}$ and 67\,Kkms$^{-1}$kpc$^{2}$ in NGC4945 and Circinus.
%
%

\citet{Curran2001,Curran1998} and \citet{Mauersberger1996} mapped the low-$J$
2--1, and 3--2 transitions of CO in the centers of both galaxies with
SEST. We smoothed these data to the resolution of the NANTEN2 data,
i.e. to $38''$. Only the $^{12}$CO 3--2 spectrum in Circinus is shown
at its original resolution of $15''$ because a map of the central
region could not be retrieved. Calibration of the CO 3--2 spectrum in
NGC\,4945 was confirmed recently at APEX \citep{Risacher2006}.  The
$^{12}$CO 1--0, $^{13}$CO 1--0 and $^{13}$CO 2--1 spectra of the central
region of NGC\,4945 and Circinus are presented in \citet{Curran2001}. 
We list integrated intensities in Table\,\ref{tab_intint_circinus}.

NGC\,4945 shows broad emission between 350 and 800\,kms$^{-1}$.  The
CO 4--3 line shape resembles the line shapes of 1--0 and 2--1.The
line shape of \CI\ 1--0 is similar to that of the CO transitions with a slighlty
higher peak temperature than CO 4--3. 
%

In CO 1--0 and 2--1, Circinus shows broad emission between 200 and
about 600\,kms$^{-1}$. The velocity component at $\sim550$\,kms$^{-1}$
becomes weaker with rising rotational number. Emission of \CI\ 1--0 is
restricted to 200 and $\sim500$\,kms$^{-1}$ only.
In Circinus, the \CI\ peak temperature  is a factor $\sim3$ stronger than CO 4--3.

%

\begin{center}   
\begin{table}[h*]   
\caption[]{\label{tab_intint_circinus}   
{\small Observed line intensities in Circinus and NGC\,4945. Calibration
  uncertainties are estimated to be 15$\%$. The spectra in Figure\,\ref{fig_spectra} and the
  integrated intensities correspond to the listed angular resolution (FWHM).
References:    
$^a$\citet{Curran2001},$^b$ \citet{Mauersberger1996}.
}}   
\begin{tabular}{lrrrc  }   
\hline \hline   
 &    Circinus  & NGC\,4945 & \\  
line transition     & $I_{\rm int}$ &  $I_{\rm int}$ & FWHM  \\
     & [Kkms$^{-1}$] &  [Kkms$^{-1}$] & [$''$]  \\
\hline 
CO 1--0$^{a}$ & 180  & 510 & 45  \\   
CO 2--1$^{a}$ & 177  & 390 & 38 \\   
CO 3--2$^{b}$ & - & 330 & 38  \\   
CO 3--2$^{b}$ & 230 & - & 15  \\   
CO 4--3 & 58 & 212 & 38 & \\   
$^{13}$CO 1--0$^{a}$ & 12 & 30 & 45 \\
$^{13}$CO 2--1$^{a}$ & 19 & 45 & 38  \\
$[$CI$]$ 1--0 & 163 & 248 & 38 &  \\

%
\noalign{\smallskip} \hline \noalign{\smallskip}   
\end{tabular}   
\end{table}   
\end{center}

\section{Physical conditions} 
\subsection{LTE}

In the optically thin limit, the integrated intensities of \CI\ and
$^{13}$CO listed in Table\,\ref{tab_intint_circinus} are proportional
to the total column densities. LTE column densities of carbon are
rather independent of the assumed excitation temperatures 
\citep[e.g.][]{Frerking1989}. We find
%
%
$N_{\rm C}=3.4-3.9\, 10^{18}$cm$^{-2}$ in NGC\,4945 and
$N_{\rm C}=2.2-2.5\,10^{18}$cm$^{-2}$ in Circinus for a
temperature range of $T_{\rm ex}= 20 - 150$\,K.

We used the $^{13}$CO $J$=1--0 and $J$=2--1 integrated intensities to
derive total CO column densities, assuming LTE, optically thin
$^{13}$CO emission, a CO/$^{13}$CO abundance ratio of 40 \citep{Curran2001}, and $T_{\rm ex}=
20$\,K. We find a total CO column density of $N_{\rm CO}$=
1.0-1.7\,10$^{18}$cm$^{-2}$  and  $N_{\rm CO}$=
4.1-6.7\,10$^{17}$cm$^{-2}$ in NGC\,4945 and Circinus respectively depending
on which transition is used, 1--0 or 2--1.
Varying the temperature  to $T_{\rm ex}=150$\,K the column density of CO 
slightly increases up to $N_{\rm CO}$= 3.2\,10$^{18}$cm$^{-2}$ and 
$N_{\rm CO}$= 1.4\,10$^{18}$cm$^{-2}$ for NGC4945 and Circinus.

Another method uses the Galactic CO 1--0 to H$_{\rm 2}$ conversion
factor $X_{\rm MW}=2.3\,10^{20}$cm$^{-2}$(Kkms$^{-1})^{-1}$
\citep{Strong1988,strong_mattox1996} and the canonical CO to H$_{\rm 2}$ abundance
of $8.5\, 10^{-5}$ \citep{frerking1982} to derive  $N_{\rm CO}$=
3.5\,10$^{18}$cm$^{-2}$ for Circinus, i.e. a factor of $\sim$ 10 larger than
the LTE estimate indicating that the X-factor is only 1/10 Galactic. As the abundance of CO maybe different
in NGC4945 and Circinus this result has
to be taken with caution.


For NGC4945 \citet{Wang2004} derive an X-factor 7 times smaller than the Galactic
value, which leads to $N_{\rm CO}$= 1.4\,10$^{18}$cm$^{-2}$,
in good agreement with the LTE approximation from $^{13}$CO.  The LTE
column density derived from $^{13}$CO in Circinus also indicates an
X-factor around 10 times smaller than the Galactic value.

The CO/C abundance ratio is 0.29-0.50 in NGC\,4945 and 0.19-0.27 in Circinus
using the LTE column densities.

\subsection{Radiative transfer analysis of CO and $^{13}$CO}

We modeled the $^{12}$CO and $^{13}$CO emission lines using an escape
probability radiative transfer model for spherical clumps
\citep{Stutzki1985} using the CO collision rates of \citet{schinke1985}. This non-LTE model assumes a uniform density and
temperature in a homogenous clump. The physical parameters kinetic
temperature $T_{\rm kin}$, molecular density $n(\rm H_{2})$, and column
density $N_{\rm CO}$ determine the excitation conditions in this
model (Table\ref{tab_results}). 

In NGC\,4945 and Circinus, we used the ratios of the observed
integrated intensities of CO 1--0 to 4--3 and the $^{13}$CO 1--0 and
2--1 transitions \citep{Curran2001} to obtain column densities,
density and kinetic temperature, assuming a constant
$^{12}$CO/$^{13}$CO abundance ratio. We used an abundance
ratio of 40 in both sources in accordance with \citet{Curran2001}.
The escape probility code uses an internal clump line width $\Delta v_{\rm
  mod}$ which hardly effects the outcome of the model in the reasonable
range of 1-20kms$^{-1}$.

In a simultaneous fit a $\chi^{2}$-fitting routine then compared the $J$ line
ratios of the model output $R^{j}_{\rm mod}$ to the ratios of the
observed integrated intensities $R^{j}_{\rm obs}$ and determined the
model with the minimal $\chi^{2}$. We compute the normalized $\chi^{2}$ with
the degrees of freedom $d=J-p$ and $J$ being the number of independent ratios
 and $p$ the number of parameters, in our case $T_{\rm kin}$, $n(\rm H_{2})$ and $N_{\rm CO}$,
to be determined :

\begin{equation}
  \chi^2 = \frac{1}{d} \sum_{j=1}^{J}(R^j_{\rm mod} - R^j_{\rm obs})/\sigma_{j}.
\end{equation}

The errors $\sigma_{j}$ due to calibration uncertainties are estimated
to be 20$\%$.


In Circinus the $^{12}$CO 3--2 line is missing, leaving one degree of
freedom compared to two in NGC\,4945. 

$T_{\rm kin}$ and $n(\rm H_{2})$ are determined with this step.
To compare the modeled integrated intensities $I_{\rm mod}$ to the
absolute observed intensities $I_{\rm obs}$ we have to account for the
velocity filling, due to the velocity width $\Delta$v$_{\rm mod}$ of an
individual clump to the width of the galaxy spectrum  $\Delta$v$_{\rm obs}$ and the beam
dilution, due to the size of the modeled clump A$_{\rm cl}$ compared
to the beam area A$_{\rm beam}$.

The large velocity width of the observed spectra implies several clumps in the
beam $N_{\rm cl}$=$n\,\Delta$v$_{\rm obs}$/$\Delta$v$_{\rm mod}$ with $n\geq
1$ (Table\,\ref{tab_results}). The beam dilution
is determined by the fraction of modeled clump area to the beam size
which we express in terms of an area filling factor per clump $\phi_{\rm
  A,cl}$= A$_{\rm cl}$/A$_{\rm beam}$.
The total area filling factor is $\phi_{\rm  A}$= $N_{cl}\,\phi_{\rm  A,cl}$. 
The size of the clump with radius $R$, A$_{\rm cl}=\pi R^2$, can be
inferred via the mass $M$ and density $n(\rm H_{2})$ of the clump: $R=(3/(4\pi)
M/n)^{(1/3)}$. In summary, the modeled intensities of the individual
clumps $I_{\rm mod}$ are converted to the intensities of a clump
ensemble $I_{\rm ens}$ which can then be compared with the observed
intensities:

\begin{equation}   
 I_{\rm ens} = I_{\rm mod} \times N_{\rm cl}\times \phi_{\rm A,cl}=I_{\rm mod} \times \phi_{\rm A}.
\end{equation}   

\begin{center}   
\begin{table}[h*]   
  \caption[]{\label{tab_results} {\small Escape probability model
  results with two representative solutions for each source. $N_{\rm
  CO}$ and $N_{\rm H_{2}}$  denote the total beam
  averaged column density  of CO and H$_{2}$, i.e. local clump column
  densities weighted by the total area
  filling factor $\phi_{\rm  A}$}. $M$ denotes the total mass. $N_{\rm cl}$ is
  the number of clumps in the beam, $\Delta v_{\rm mod}$ is the modeled velocity
  width and $\phi_{\rm A}$ is the total filling factor to convert from modeled
  to observed intensities. The total cooling intensities are given in units of
  10$^{-5}$ergs$^{-1}$cm$^{-2}$sr$^{-1}$. }
\begin{tabular}{lcccc}   
\hline \hline   
 &   \multicolumn{2}{c} {Circinus} & \multicolumn{2}{c} {NGC\,4945}  \\  
\hline 
$\chi^{2}$     & 2.0 & 9.6 &  4.8 & 12.4  \\
$n(\rm H_{2})_{\rm loc}$\,[cm$^{-3}$]    & $ 10^{4}$& $10^{3}$   & $3\,10^{4}$ & $10^{3}$  \\
$T_{\rm kin}$ \,[K] & 20 & 100  & 20 & 100   \\
$N_{\rm CO}$ \,[10$^{16}$cm$^{-2}]$ & 35 & 50  & 76 & 63   \\
$N_{\rm C}$ \,[10$^{16}$cm$^{-2}]$ & 230 & 30  & 330 & 98   \\
$N_{\rm H_{2}}$ \,[10$^{20}$cm$^{-2}]$ &  37 & 46.5 & 89 & 74    \\
$M$ [10$^{6}$\msun] & 630 & 792  & 1385 & 1114   \\
$\Delta v_{\rm mod}$ \, [kms$^{-1}$] &  5 & 5 & 10 & 10   \\
$\Delta v_{\rm obs}$ \, [kms$^{-1}$] &  186 & 186 & 188 & 188   \\
$N_{\rm cl}$ & 50 & 38 & 35 & 40     \\
$\phi_{\rm A}$ & 2.0 & 6.3 & 1.5 & 4.0    \\
$^{12}$CO/$^{13}$CO abundance ratio & 40 & 40 & 40  & 40     \\
CO/C abundance ratio & 0.15 & 1.67 & 0.23  & 0.64     \\
\CI\ cooling intensity  & 4.1 & 7.88 & 7.2  & 11.8\\
CO cooling intensity  & 2.1 & 2.8 &  6.6 & 5.7 \\
\CI\ /CO cooling intensity ratio & 2.0 & 2.8 & 1.1 & 2.1    \\

%
\noalign{\smallskip} \hline \noalign{\smallskip}   
\end{tabular}   
\end{table}   
\end{center}

\subsubsection{NGC\,4945}

\begin{figure}[h]   
  \centering   
  \includegraphics[angle=-90,width=8cm]{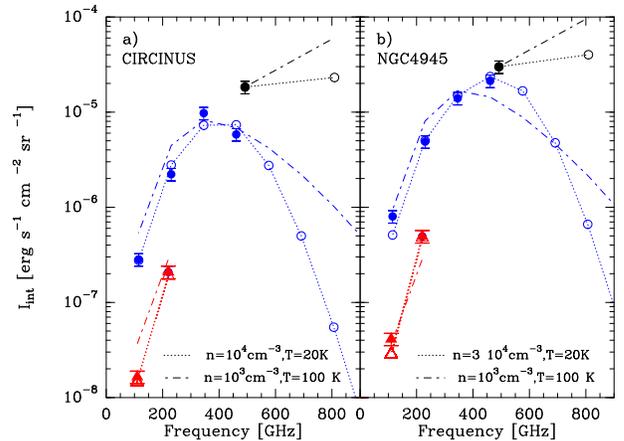}   
\caption{Radiative transfer modeling results and observations: Filled blue points show the observed CO and blue circles
  the modeled CO. The dotted line indicates the best fit solution. Filled red
  triangles show the observed $^{13}$CO and red
  triangles the modeled $^{13}$CO. Dash-dotted lines show a higher temperature
  solution. Filled black points are the observed $[$CI$]$ integrated
  intensities and black circles the predicted $[$CI$]$ integrated
  intensities.}  
\label{fig_seds}   
\end{figure}

Figure\,\ref{fig_seds}b shows the observed intensities of CO,
$^{13}$CO, and \CI\ together with two representative solutions of the
radiative transfer calculations (see also Table\,\ref{tab_results}).
%

\paragraph{CO.}

Fitting the CO and $^{13}$CO lines, we find a degeneracy in the
n(H$_{2})$-T$_{\rm kin}$ plane of the solutions for a rather constant
pressure n(H$_{2}) \times \, T_{\rm kin}$ $\sim 10^{5}$Kcm$^{-3}$.
The best fits are achieved for a $^{12}$CO/$^{13}$CO abundance ratio
of 40, similar to the value found by \citet{Curran2001}.
Low $\chi^{2}$-values (see Table\,\ref{tab_results}) constrain the densities to $n(\rm H_{2})=10^{3}-10^{5}$cm$^{-3}$ and temperatures to a wide range of
$T_{\rm kin}$=20-180K with higher temperature solutions corresponding to lower
densities. 
%
%
The best fitting solution is $n(\rm H_{2})=3\,10^4$cm$^{-3}$ and $T_{\rm
  kin}$=20K. However, this solution is not significantly better than
e.g. $n(\rm H_{2})=10^3$cm$^{-3}$ and $T_{\rm kin}$=100K
(Fig.\ref{fig_seds}b, Table\,\ref{tab_results}).

The fitted column density $N_{\rm CO}$ agrees with the LTE
approximation for the CO-column density within a factor of 2--3. 
%

The peak of the modeled CO cooling curve at $J$=4 contains 35.6\% of the total
$^{12}$CO cooling intensity of $6.6\,10^{-5}$
ergs$^{-1}$cm$^{-2}$sr$^{-1}$ computed by summing the cooling intensity
of the $^{12}$CO transitions from $J$=1 to 20 for the $T_{\rm
  kin}$=20K solution.  

We use the radiative transfer model to predict the $^{12}$CO 7--6
intensity. It is rather weak, depending strongly on $T_{\rm kin}$. It varies
from k 1.3 Kkms$^{-1}$  
 for the $T_{\rm
 kin}$=20K solution to 4.1 Kkms$^{-1}$ for the $T_{\rm kin}$=100K fit.
  
\citet{Curran2001} find in this source $n(\rm H_{2})$=3$\times
10^{3}$cm$^{-3}$ and $T_{\rm kin}$=100K from $^{12}$CO observations of
the 3 lowest transitions and $^{13}$CO data of the two lowest
transitions. In their multi-transition study \citet{Wang2004} estimate
a density of $n(\rm H_{2})$= $10^{3}$cm$^{-3}$ for an assumed
temperature $T_{\rm kin}$=50K from CO transitions up to $J=3$. In contrast, 
the observed CN and CH$_{3}$OH lines indicate local
densities around $10^4$cm$^{-3}$. The solutions for CO and $^{13}$CO found in
the literature
are in good agreement with our parameter space of solutions showing
that the additional CO 4--3 line does not help significantly to
improve the fits. 

\paragraph{Atomic carbon.}

Assuming the same density, kinetic temperature, velocity filling, and
beam dilution for carbon as for CO, we use the observed \CI\ intensity
and the radiative transfer model to estimate the carbon column density
and hence the CO/C abundance ratio. The CO/C abundance ratio varies between 0.23
and 0.64 for the two solutions listed in Table\,\ref{tab_results}.The
corresponding carbon column densities are $N_{\rm  C}=3.3\,10^{18}$cm$^{-2}$
and $N_{\rm  C}=9.8\,10^{17}$cm$^{-2}$ for the high and the low density
solution, respectivly. The column density for the latter solution is a factor of 3
lower compared to the LTE carbon column density. The assumption of optically
thin emission in the LTE might be overestimating the column density compared
to the escape probability modelling.
%
%
We also use the model to predict the \CI\ 2--1 intensity. It is 16
Kkms$^{-1}$ for the $T_{\rm kin}$=20\,K/$n(\rm H_{2})=3\,10^4$\,cm$^{-3}$ solution.
However, this result critically depends on the kinetic
temperature. The $T_{\rm kin}$=100\,K/$n(\rm H_{2})=10^3$\,cm$^{-3}$ solution
yields 182\,Kkms$^{-1}$. Also, the \CI\ 2--1/\CI\ 1--0 line ratios
change from 0.07 to 0.75, depending on the solutio.This shows that the high-lying
transitions can be used to break the degeneracy.

The total cooling intensity of both \CI\ lines is listed in Table\,\ref{tab_results} for both
presented solutions. The cooling intensity of the two \CI\ lines is thus of the order of the total
cooling intensity of CO with the C/CO cooling intensity ratio varying between 1.1-2.1
for the discussed solutions.

\subsubsection{Circinus}

\paragraph{CO.}

The modeled CO cooling curves are shown in Figure\,\ref{fig_seds}a and
the fit results are summarized in Table\,\ref{tab_results}.The CO line ratios
are given for a $\Delta\,v_{mod} $=5kms$^{-1}$ so there are about 40 clouds in the beam to account for the observed
velocity width of $\Delta\,v_{obs}$=186kms$^{-1}$.  The assumed
$^{12}$CO/$^{13}$CO abundance ratio of 40 is slightly lower than the values
of $\sim$ 60, found by \citet{Curran2001}.


Good fits corresponding to low $\chi^{2}$ (see Table\,\ref{tab_results}) can be found for densities
of $n(\rm H_{2})=10^{3}-10^{4.5}$cm$^{-3}$ and a large range of
temperatures of $T_{\rm kin}$=20-160K while the product of $n(\rm H_{2}) \times \, T_{\rm kin}$ stays approximately constant at $\sim
10^{5}$Kcm$^{-3}$. Again, a number of solutions provide
consistent CO cooling curves.

The lowest $\chi^{2}$ is obtained for $n(\rm H_{2})=10^{4}$cm$^{-3}$,
$T_{\rm kin}$=20K and a column density of $N_{\rm
  CO}=3.5\,10^{17}$cm$^{-2}$ assuming 50 modeled clumps in the beam.
The column density $N_{\rm CO}$ is well determined showing a steep
gradient of $\chi^{2}$-values for varying densities and temperatures.
This is in reasonable agreement with the LTE-approximation from $^{13}$CO. 
A second solution at $n=10^{3}$cm$^{-3}$ and $T_{\rm kin}$=100K also lies
within the 1$\sigma$-contour of the $\chi^2$ distribution (cf.
Fig.\ref{fig_seds} and Tab.\ref{tab_results}).

The results agree well with the solutions found by \citet{Curran2001}.
They find $T_{\rm kin}$=50-80K and $n(\rm H_{2})=2\,10^{3}$cm$^{-3}$
from observations of the 3 lowest $^{12}$CO transitions and the 2
lowest $^{13}$CO transitions.

The modeled CO cooling curve peaks at $J$=4 containing 35\% of the total
$^{12}$CO cooling intensity of $2.1\,10^{-5}$
ergs$^{-1}$cm$^{-2}$sr$^{-1}$ for the $T_{\rm kin}$=20K fit. 

The predicted integrated intensity of $^{12}$CO 7--6 is very weak, varying strongly from
0.1 Kkms$^{-1}$ for the $T_{\rm kin}$=20K solution to
1.9 Kkms$^{-1}$ for $T_{\rm kin}=100$ K.    


\paragraph{Atomic carbon.}

Assuming the same density and temperature for carbon as for CO, we use
the best fitting CO model to derive the \CI\ 1--0 intensity, carbon
column densities, and CO/C abundance ratios.  The predicted CO/C
abundance ratio is 0.15, again consistent with the optically thin LTE
result. The predicted \CI\ 2--1 integrated intensity is 6.1
Kkms$^{-1}$.

As in NGC\,4945, changing the temperature has a large effect on the \CI\
2--1 intensity. The $T_{\rm kin}$=100K solution yields 111 Kkms$^{-1}$
and a higher CO/C abundance ratio of 1.67. Thus the \CI 2--1/\CI 1--0 line
ratios changes from 0.04 to 0.74.

The corresponding carbon column densities for the two
  presented solutions are $N_{\rm  C}=2.3\,10^{18}$cm$^{-2}$
and $N_{\rm  C}=3.0\,10^{17}$cm$^{-2}$, repectivly. For the latter solution the column
density is about a factor of 10 lower compared to the LTE carbon column density.
The optically thin assumption for the LTE modelling is obviously not appropriate in
the high temperature and low density scenario.

The total cooling intensity ratio of \CI\/CO varies from 2.1 to 2.8 for
the presented solutions.Carbon is a stronger coolant than CO by a factor of 2-3.

\section{Discussion}
\subsection{\CI\ 1--0 luminosities}

The \CI\ 1--0 luminosities for the centers of the Seyfert galaxies NGC\,4945
and Circinus are 91 and 67\,Kkms$^{-1}$kpc$^{2}$ (Fig.\,\ref{fig_israel}).
To date, about 30 galactic nuclei have been studied in the 1--0 line
of atomic carbon, most of which are presented in \citet{Gerin2000} and
\citet{IsraelBaas2002}. The \CI\ luminosity of a
source is an important property since it gives the amount of energy
emitted per time which is proportional to the number of emitting
atoms, i.e. proportional to the \CI\ column density in the limit of
optically thin emission. The NANTEN2 38\arcsec\, beam achieves
$\sim$700\,pc resolution in Circinus and NGC\,4945 which both lie at
$\sim$4\,Mpc. To achieve the same spatial resolution for galaxies at
$\sim$12\,Mpc distance, e.g.  for NGC\,278, NGC\,660, NGC\,1068,
NGC\,3079 and NGC\,7331 listed in \citet{IsraelBaas2002}, one would
need an angular resolution of $\sim13''$, comparable to the $10''$
JCMT beam at 492\,GHz. The luminosities studied in these 7 sources
thus all sample the innermost $\sim$700\,pc.  Area integrated \CI\
luminosities are found to vary strongly between $\sim$1 and
$\sim160$\,Kkms$^{-1}$kpc$^{2}$ in these 7 galaxies (Fig.\,\ref{fig_israel})
\citep{Israel2005,IsraelBaas2002}.  Quiescent centers show modest
luminosities $1\le L$(\CI)$\le5$\,Kkms$^{-1}$kpc$^{2}$, while starburst
nuclei in general show higher luminosities. The largest luminosities
are found in the active nuclei of NGC\,1068 and NGC\,3079 which show
50 and 160\,Kkms$^{-1}$kpc$^{2}$ \citep{IsraelBaas2002}. NGC4945 and Circinus
also fall in this category (Fig.\,\ref{fig_israel}).
\subsection{\CI\ 1--0/CO 4--3 line ratios}

\begin{figure}[h]   
  \centering   
  \includegraphics[angle=-90,width=8cm]{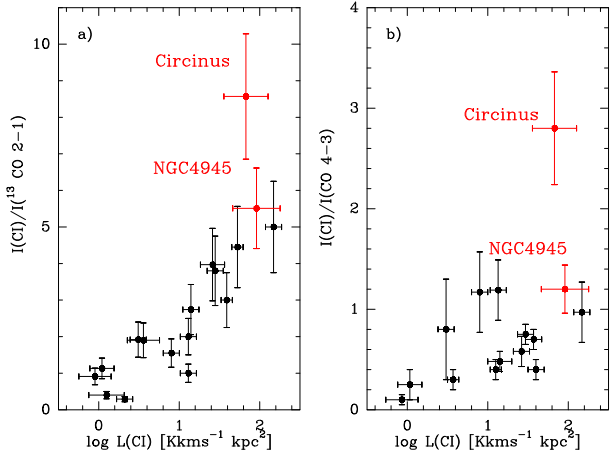}   
\caption{a \CI 1-0/$^{13}$CO 2--1 ratios versus L$(\CI)$. b)
 \CI 1-0/CO 4--3 ratios versus L$(\CI)$. Black points are the results
 from \citet{IsraelBaas2002} and red points are the ratios from this paper.}  
 
\label{fig_israel}   
\end{figure}

The \CI\ 1--0/CO 4--3 ratio of integrated intensities is 1.2 in
NGC\,4945 and 2.8 in Circinus. For Circinus the ratio is larger than any ratios
previously observed in other galactic nuclei or in the Milky Way.
The \CI 1-0/CO 4--3 ratio is shown in
Figure\,\ref{fig_israel} versus the area integrated \CI\, 
luminosity. As also discussed in \citet{IsraelBaas2002}, we see no functional
dependence. Galactic sources (not shown in this figure) would lie in
the lower left corner.
\citet{Israel2005} studied 13 galactic nuclei and found that the \CI\
1--0 line is in general weaker than the CO 4--3 line, but not by much.
Ratios vary over one order of magnitude between 0.1 in Maffei2 and 1.2
NGC\,4826.  Galactic star forming regions like W3\,Main or the Carina
clouds show much lower values, between about 0.1 and 0.5, which are
consistent with emission from photon dominated regions (PDRs)
\citep{Kramer2004,Jakob2007,Kramer2007}. \citet{Fixsen1999} find 0.22 in the
Galactic center and 0.31 in the Inner Galaxy. The variation seen in the
various Galactic and extragalactic sources appears to be intrinsic and
not due to observations of different angular resolutions. This is
because the frequencies of the two lines are very close and angular
resolutions are therefore similar if the same telescope is used.

\subsection{\CI\ 1--0/ $^{13}$CO 2--1 line ratios}
The \CI\ 1--0/ $^{13}$CO 2--1 line ratios in NGC\,4945 and Circinus are 5.51 and
8.57, respectively (Fig.\,\ref{fig_israel}).In NGC4945, the
observed ratio is consistent with results found in previous studies
\citep{IsraelBaas2002,Gerin2000} ranging up to ratios of 5. For Circinus,
the ratio is again higher than any previous measurement.  

Two thirds of the sample of galaxies studied by \citet{IsraelBaas2002} show \CI\ 1--0/ $^{13}$CO 2--1
line ratios well above unity. The sample consists of quiescent, star burst  
and active nuclei. The highest \CI\ 1--0/ $^{13}$CO 2--1 ratios are found
in star burst and active nuclei, consistent with our
observations. \citet{Gerin2000}  find a similar result with
two thirds of the galaxies in their sample exceeding a ratio of 2.
High ratios can be qualitatively understood in low column density enviroments
with mild UV radiation fields. In these regimes most CO will be dissociated
and the gas-phase carbon will be neutral atomic \citep{IsraelBaas2002}.  
This implies that dense, star forming molecular cloud cores are not the major
emission source in galaxy centers.

In general the studied centers of external galaxies show stronger \CI\
emission than one would expect from Galactic observations, which show typical
ratios of 0.2-1.1 \citep{Mookerjea2006}. In Galactic sources high ratios are
found in low gas column densitis and medium UV
radiation enviroments where $^{13}$CO will be dissociated and atomic carbon
remains neutral in the gas phase i.e. in translucent clouds and at cloud
edges \citep{Israel2005}. He concludes that
the dominant emission from galaxy centers does not stem from PDRs.

\citet{Meijerink2007} studied irradiated dense gas in galaxy nuclei using a
grid of XDR and PDR models. For the same density the predicted \CI\ 1--0/ $^{13}$CO 2--1 line
ratios are significantly higher for the XDR- compared to PDR-models 
(Fig. 10 in Meijerink et al. 2007). We observed ratios of 61 and 95 in NGC4945
and Circinus respectivly, on the erg-scale.  
These ratios can be explained by XDR-models at high densities $n(\rm
H_{2})=2\,10^{3}-10^{5}$cm$^{-3}$. PDR-models explain the observed ratios in low
density regimes with $n(\rm H_{2})=2\,10^{2}-6\,10^{2}$cm$^{-3}$. The high \CI\
1--0/ $^{13}$CO 2--1 line ratios observed in NGC\,4945 and Circinus may hint at
a significant role of X-ray heating in these galaxy nuclei as our predicted
densities of $n(\rm H_{2})=10^{3}-10^{4}$cm$^{-3}$ agree with the high-density
XDR-models in \citet{Meijerink2007}.
\\
\subsection{Total CO and \CI\ cooling intensity}

In all sources studied by \citet{bayet2006} but NGC\,6946, the CO
cooling intensity exceeds that of atomic carbon. They find the \CI\ to CO
cooling ratio to vary between $\sim0.3$ for M83 to
$\sim 2$ for NGC\,6946. NGC\,4945 and Circinus show similarly high values as
the latter galaxy:
In NGC\,4945 this ratio is $\sim 1-2$ and in
Circinus we find $\sim 2-3$.  These values are around 2 orders of
magnitude higher than the typical values found in Galactic star
forming regions \citep{Jakob2007,Kramer2007}.

\subsection{Shape of the CO cooling curve}

The modeled CO cooling curve of NGC\,4945 and Circinus peaks at
$J=4$. Observations of the higher lying CO lines i.e. the CO 6-5 and CO 7-6 lines
will however be important to verify our model predictions and the importance of CO
cooling relative to C.

The shape and maximum of the cooling curve of $^{12}$CO has been studied in a number of nearby and
high-z galaxies. \citet{Fixsen1999} find a peak
at $J$=5 in the central part of the Milky Way using FIRAS/COBE data
and rotational transitions up to 8--7. \citet{bayet2006} observed 13
nuclei in mid-$J$ CO lines up to 7--6 and 
find that the peak of the cooling curves vary with nuclear
activity. While normal nuclei exhibit peaks near $J_{\rm up}=4$ or 5,
active nuclei show a peak near $J_{\rm up}=6$ or 7. NGC\,253 was
observed at APEX in CO upto $J_{\rm up}=7$ and shows a maximum at
$J=6$ \citep{Guesten2006}.
%
%
On the other hand, studies of high redshift galaxies show cooling
curves peaking as low as $J$=4 for SMM\,16359 \citep{WeissA2005}, as
high as $J$=9 in the case of the QSO APM\,08279 \citep{WeissA2007}, and
peaking at $J$=7 for the very high redshift ($z=6.4$) QSO J1148
\citep{WalterF2003}.
%

\subsection{Pressure of the molecular gas}

A wide range of temperatures $T_{\rm kin}= 25-150$K and densities
$n(\rm H_{2})=5\,10^{2}- 7\,10^{5}$cm$^{-3}$ has been found in
similar studies of external nuclei, including ULIRGs, normal spirals,
star burst galaxies, and interacting galaxies
\citep[e.g.][]{bayet2006,IsraelBaas2002}. In the irregular galaxy
IC\,10 rather high densities $n(\rm H_{2})\sim 10^6$cm$^{-3}$ and low
temperatures of $T_{\rm kin}= 25$K are found while the molecular
  gas in the center of the spiral galaxy NGC\,6946 is found to be
less dense, $n(\rm H_{2})=10^{3}$cm$^{-3}$, but much hotter, $T_{\rm
  kin}=130$K.  \citet{Guesten2006} studied NGC\,253 to obtain
 $n(\rm H_{2})=10^{3.9}$cm$^{-3}$ and $T_{\rm kin}=60$K while
\citet{bradford2003} investigated the same source and derived a higher
$T_{\rm kin}=120$K and density $n(\rm H_{2})=4.5 \, 10^{4}$cm$^{-3}$.
Both studies used $^{12}$CO 7--6 observations.
However, the temperature/density degeneracy cannot be resolved.  The
solutions we present in our study of NGC\,4945 and Circinus show
densities and temperatures of $n(\rm H_{2})=10^{3}-10^{4}$cm$^{-3}$ with a less
well constrained temperature $T_{\rm kin}=20-100$K; depending on the
density, as discussed before.

\subsection{\CI\ 2--1/1--0 line ratio}

The modeled \CI\ 2--1/1--0 line ratios change from 0.07 to 0.75 in
NGC\,4945 and from 0.04 to 0.74 in Circinus for the presented escape
probability solutions. Observed ratios vary from 0.48 in G333.0-0.4 \citep{Tieftrunk2001}
 to 2.9 in W3 main for galactic and extragalactic sources
\citep{Kramer2004}. For M82, \citet{stutzki1998} found a ratio of
0.96. \citet{bayet2006} observed ratios ranging between
1.2 (in NGC\,253) to 3.2 (in IC\,342). The \CI\ 2--1/1--0 line ratio for
the low temperature solution predicted for NGC4945 and Circinus is
significantly lower than previous results in
the literature.

\subsection{CO/C abundance ratio}


Compared to an abundance ratio CO/C of 3-5 in NGC\,253
\citep{bayet2004} and an average value of 2 in the nucleus of M83
\citep{White1994,IsraelBaas2001}, NGC\,4945 and Circinus have a higher
fraction of atomic carbon in their nuclei which resembles the low CO/C
abundance ratios found in Galactic translucent clouds.  We find values of
0.23-0.64 in NGC\,4945 and 0.15-1.67 in Circinus. Due to the degeneracy of $n(\rm H_{2})$,T
the abundance ratios cannot be determined more accurately. In galactic molecular
clouds values range between 0.16-100, low values are found in
translucent clouds \citep{Stark1994} while massive star forming
regions show high CO/C abundances \citep{Mookerjea2006}.

\subsection{Future observations}

Future observations of CO 7--6 and \CI\ 2--1 will be important to 
better understand the the kinetic temperature and density to resolve their
degeneracy, and for an understanding of the dominating heating mechanism
i.e.  X-ray or UV heating. We estimate detection of the \CI\ 2--1 line
with NANTEN2 within less than 10 minutes total observation time
under average weather conditions . The CO 7-6 will be harder to detect,
depending on the excitation conditions. At 810 GHz the 1\,GHz
bandwidth of the receiver translates into a velocity range of
370kms$^{-1}$. We therefore plan to stack future observations to be
able to cover the broad line widths of Circinus and NGC\,4945.

\begin{acknowledgements}   
  We thank S.\,Curran for providing us with the SEST-data for Circinus,
  R. Mauersberger and G. Rydberg for SEST-data for NGC\,4945,  and
  R.G\"usten and S. Philipp for the APEX-data of NGC\,253.
 The NANTEN2 project (southern submillimeter observatory consisting
 of a 4-meter telescope) is based on a mutual agreement between
 Nagoya University and The University of Chile and includes member
 universities from six countries, Australia, Republic of Chile,
 Federal Republic of Germany, Japan, Republic of Korea, and Swiss
 Confederation. We acknowledge that this project could be realized by
 financial contributions of many Japanese donators and companies.
 This work is financially supported in part by a Grant-in-Aid for
 Scientific Research from the Ministry of Education, Culture, Sports,
 Science and Technology of Japan (No.15071203) and from JSPS (No.
 14102003 and No.18684003), and by the JSPS core-to-core program
 (No.17004).
  This work is also financially supported in part by the grant
  SFB\,494 of the Deutsche Forschungsgemeinschaft, the Ministerium
  f\"ur Innovation, Wissenschaft, Forschung und Technologie des Landes
  Nordrhein-Westfalen and through special grants of the Universit\"at
  zu K\"oln and Universit\"at Bonn.
  L. B. and J. M. acknowledge support from the Chilean Center for
  Astrophysics FONDAP 15010003.   
\end{acknowledgements}   
\bibliographystyle{aa} 
\bibliography{p_nanten_galaxies} 
   
\clearpage 
\end{document}